\documentclass[reprint, showpacs,floatfix,twocolumn,10pt]{revtex4-1}

\usepackage{graphicx}
\usepackage{amsmath}
\usepackage{amsfonts}
\usepackage{amssymb}
\usepackage{amsthm}
\usepackage{dsfont}
\usepackage[caption=false]{subfig}
\usepackage{nameref}
\usepackage[colorlinks]{hyperref}
\usepackage{color}
\usepackage[utf8]{inputenc}
\usepackage[outdir=./]{epstopdf}

\newcommand{\avg}[1]{\ensuremath{\left\langle#1\right\rangle}}
\newcommand{\bra}[1]{\ensuremath{\left\langle#1\right|}}
\newcommand{\ket}[1]{\ensuremath{\left|#1\right\rangle}}
\newcommand{\ketbra}[2]{\ket{#1}\!\!\bra{#2}}

\newtheorem{theorem}{Theorem}
\newtheorem{corollary}{Corollary}[theorem]
\newtheorem{observation}{Observation}

\DeclareMathOperator{\tr}{tr}

\begin{document}

\title{How an autonomous quantum Maxwell demon can harness correlated information}

\author{Adrian Chapman}
\email{akchapman@unm.edu}
\author{Akimasa Miyake}
\email{amiyake@unm.edu}
\affiliation{Center for Quantum Information and Control, Department of Physics and Astronomy, University of New Mexico, Albuquerque, NM 87131, USA} 

\begin{abstract}
We study an autonomous quantum system which exhibits refrigeration under an information-work trade-off like a Maxwell demon. The system becomes correlated as a single ``demon" qubit interacts sequentially with memory qubits while in contact with two heat reservoirs of different temperatures. Using strong subadditivity of the von Neumann entropy, we derive a global Clausius inequality to show thermodynamic advantages from access to correlated information. It is demonstrated, in a matrix product density operator formalism, that our demon can simultaneously realize refrigeration against a thermal gradient and erasure of information from its memory, which is impossible without correlations. The phenomenon can be even enhanced by the presence of quantum coherence.
\end{abstract}

\maketitle

\section{Introduction}
\label{sec:Introduction}

With the discovery of his famous gedanken experiment, James Clerk Maxwell started physics on the long road to the unification of thermodynamics with information theory \cite{maruyama2009maxwell, jarzynski2011equalities, parrondo2015thermoandinfo}. Nearly 200 years later, with the ever-advancing miniaturization of devices \cite{burek2012freestanding, cui2001functional, gudiksen2002growth, hu2007heterostructure, maletinsky2012robust}, the idea that one could exploit detailed knowledge of a system's microstate as a thermodynamic resource no longer seems like a mere idealization \cite{broeck2012efficiency, mandal2012maxwell, hartich2014stochastic, strasberg2014secondlaws, shiraishi2015roleofmeasurement, horowitz2015multipartite, barato2015unifying}. To the contrary, we hope to someday soon push our technology to the regime where quantum effects become relevant \cite{delrio2011thermodynamic, deffner2013informationproc, jacobs2012quantum, caves1990quantitative, scully2001extracting, horodecki2005local, hilt2010validity, zurek2003quantum, oppenheim2002thermodynamical, segal2008stochastic, rudolph2015description, goold2015role}. Since there are more delicate constraints on correlations in the quantum world, like strong subadditivity of von Neumann entropy, as well as a variety of monogamy relations (exclusive trade-offs) among different kinds of correlations, it is timely to consider a question unforeseen in Maxwell's time: {\it How can correlations be harnessed for the performance of these devices in the fully general quantum setting?}

This significant question has been studied extensively from different perspectives, and it has been mentioned that correlations can yield an advantage in performing certain thermodynamic tasks, such as work extraction \cite{perarnaullobet2014extractable, delrio2014relative, funo2013thermodynamic, oppenheim2002thermodynamical, robnagel2014nanoscale,brunner2014entanglement, dillenschneider2009energetics, gelbwaserklimovsky2014power, hovhannisyan2013entanglement}. In this paper, we focus on two key challenges: (i) accounting for the contribution of correlations to the second law of thermodynamics and (ii) comparing quantum and classical correlations as thermodynamic resources on the same footing. Regarding (i), it is important to notice that the system is coupled to external degrees of freedom in quantum open-system dynamics, and the resulting correlations can be utilized, in principle, by an external, {\it classical} agent to ``measure" the system and gain knowledge in the framework of feedback control \cite{shiraishi2015roleofmeasurement}. However, this utilization of correlations should be distinguished from the way an internal agent like a quantum Maxwell demon, or a part of the quantum system, handles correlations, since the latter is the scenario we are interested in here. Regarding (ii), the majority of prior works \cite{delrio2011thermodynamic, zurek2003quantum, caves1990quantitative, lloyd1997quantum, jacobs2012quantum, perarnaullobet2014extractable, deffner2013information} defines quantum and classical Maxwell demons in such a manner that the classical version is treated as a limiting case (typically with a fixed basis). This obviously makes the classical demon at most as powerful as the quantum one, but it may not immediately imply that classical correlations are at most as useful as quantum correlations for the demon who can handle both kinds.

To address point (i), we study a quantum system which exhibits an information-work trade-off independently of any external agent. Following a pioneering work of the classical autonomous Maxwell demon studied in Ref.~\cite{mandal2012maxwell}, we consider a scenario in which a qubit, in contact with two heat reservoirs of different temperatures, accesses sequentially many memory qubits, and we analyze their correlations in the fully quantum regime. We think this to be among the simplest models to exhibit the thermodynamic features akin to a quantum Maxwell demon. For point (ii), as defined formally in the next section, we stress that our demon is {\it identical} regardless of the nature of information in its memory. This allows us to unambiguously define work performed by the demon, since there has otherwise been difficulty in doing so due to the intrinsic ``uncertainty" of quantum systems \cite{gallego2015work, gemmer2015work, aberg2013work, boukobza2006thermodynamics, goswami2013thermodynamics, horodecki2013fundamental, schroder2010work, skrzypczyk2013extracting, dahlsten2011inadequacy, woods2015maximum}. 

Here we discover a new nonequilibrium phase of our demon: refrigeration against a thermal gradient coincident with memory erasure at the expense of correlations. We find that correlations enable our demon to exploit quantum coherence to realize an advantage over its classical counterpart.  To handle the complexity of correlations, we apply a tool from condensed-matter physics---the matrix product density operator formalism---which makes tractable our calculations in the fully quantum-correlated regime. This technique exploits the fact that correlations built under local interactions change locally \cite{verstraete2004MPDO, zwolak2004mixed, eisert2013entanglement, orus2014practical}. It thus gives a powerful tool for treating correlations in the thermodynamic framework, a task which has otherwise remained formidable. We expect our proof-of-principle to inspire a cross-fertilization of quantum many-body physics and quantum information with quantum thermodynamics.

\newpage

Section \ref{sec:Background} is an introduction and overview of the autonomous quantum demon. In Sec. \ref{sec:gcinequality}, we apply techniques from quantum information theory to derive its effective second law constraint. We give an analytic solution using our matrix product density operator formalism in Sec. \ref{sec:MPDOsoln}. In Sec. \ref{sec:Numerics}, we demonstrate the existence of a phase of simultaneous refrigeration and erasure in the presence of correlations, and in Sec. \ref{sec:quantadv}, we demonstrate the quantum advantage. Finally, we close with a discussion in Sec. \ref{sec:discussion}.

\section{Overview of the Model}
\label{sec:Background} 

\begin{figure}[h!]
\subfloat[]{\label{fig:quantumdemon}
  \includegraphics[width=0.44\textwidth]{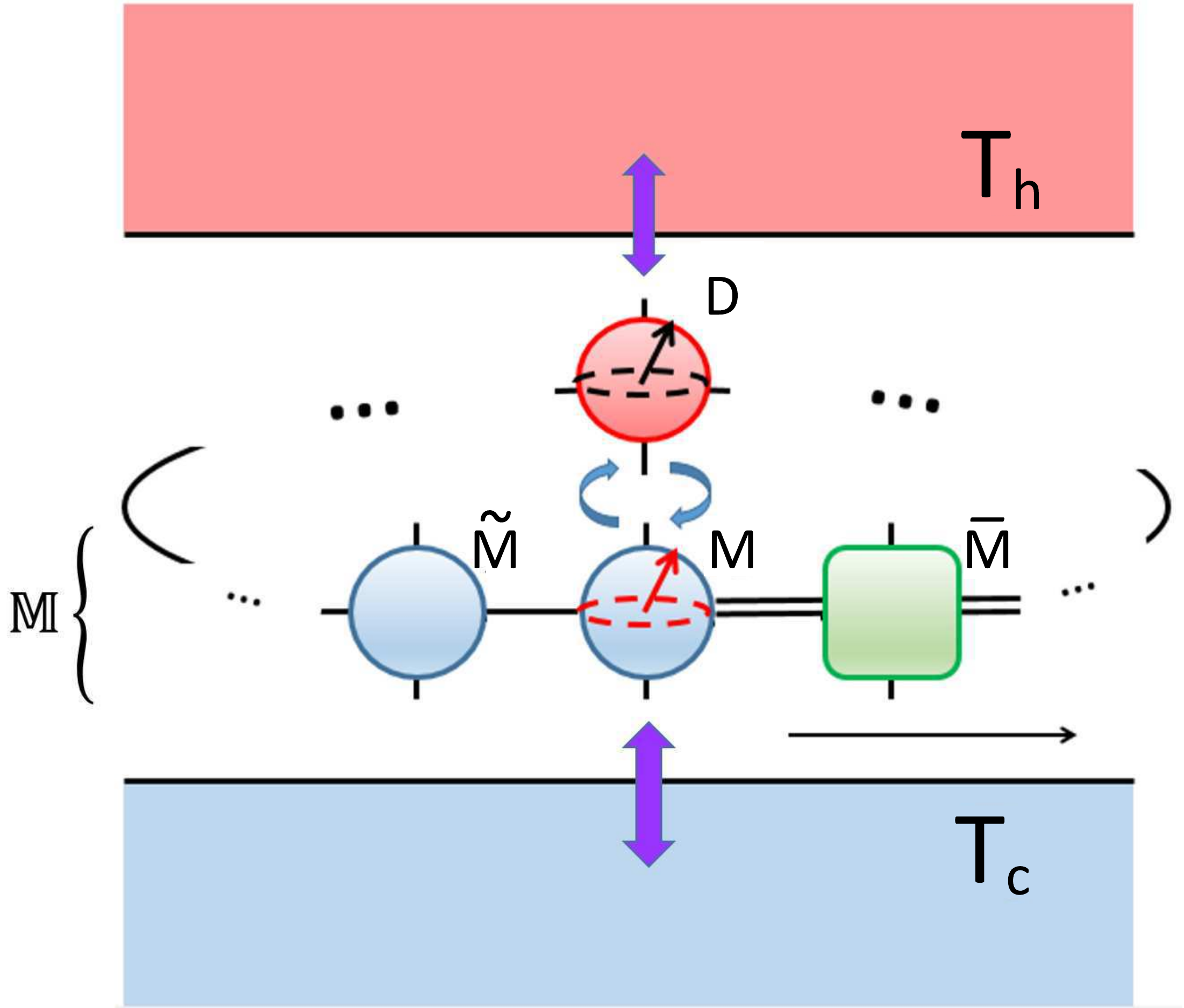}
 	}\\
\subfloat[]{\label{fig:transitions}
  \includegraphics[width=0.45\textwidth]{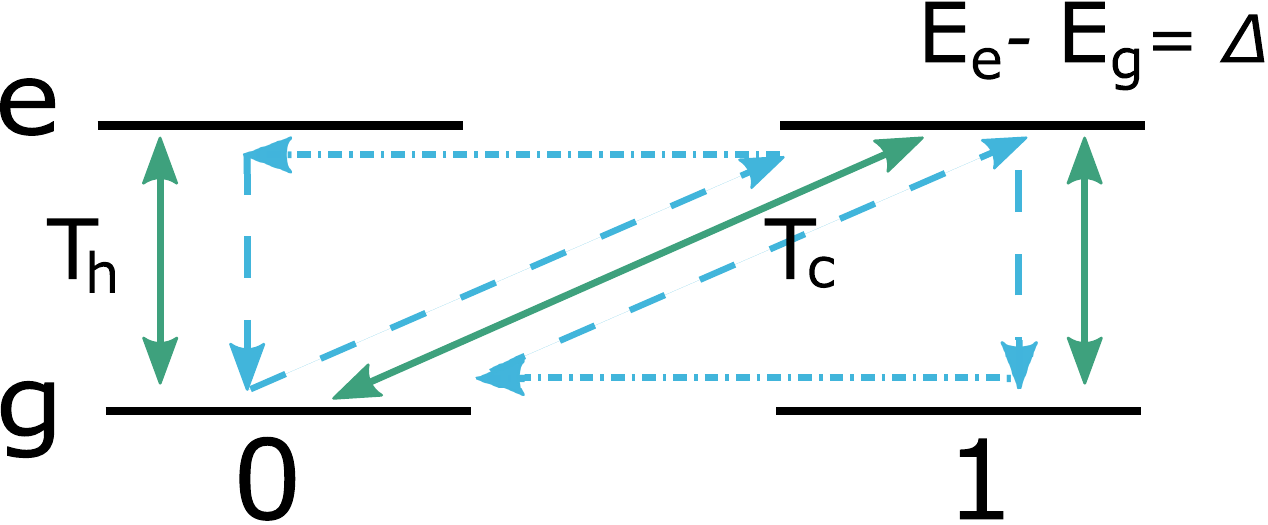}
 	}
	\caption{(Color online) (a) A snapshot of our quantum Maxwell demon. The demon qubit $D$ interacts sequentially with each qubit in a memory $\mathds{M}$ via an open-system process in contact with two thermal reservoirs of different temperatures. For a given interaction, $M$ is the currently interacting qubit, $\bar{M}$ the subsystem of previously interacting qubits, and $\tilde{M}$ that of qubits yet to interact. The joint state of $D \mathds{M} $ is described by a matrix product density operator with periodic boundary conditions, though we take sufficiently many interactions that the system has reached a periodic steady state. Interaction with the shared system $D$ allows the possibility to build further correlations within $\bar{M}$, hence the double line. (b) The transition diagram of an individual interaction between the demon qubit and a memory qubit. Solid arrows represent possible stochastic transitions, allowed by exchanging a unit $\Delta$ of energy with the associated reservoir, with transition rates satisfying the detailed balance condition. For example, intrinsic transitions coupled with the thermal reservoir of temperature $T_h$ change only the state of the demon, whereas cooperative transitions coupled with that of $T_c (< T_h)$ change the joint state of  the demon and the memory qubit, leaving a record on the memory. The dashed arrows indicate the cyclic processes which refrigerate (pump energy against the gradient), with the horizontal dot-dashed transitions provided by the memory shift. Note both refrigerative processes result in a net flip of the memory qubit from ``0" to ``1."}
\label{fig:model}
\end{figure}

Our autonomous Maxwell demon [Fig.~\ref{fig:model}(a)] is a generalization of the model proposed in Ref.~\cite{mandal2012maxwell}. The ``demon" $D$ is a qubit spanned by orthonormal basis states $\ket{g}$ and $\ket{e}$ with energy gap $E_{e} - E_{g} = \Delta$. It interacts sequentially with qubits in an infinitely long one-dimensional array $\mathds{M}$, called the memory. The qubits in $\mathds{M}$ are energetically degenerate and may be initially correlated, though they are noninteracting among themselves. The memory thus acts only as an informational, and not an energy, resource.
When discussing a given interaction, we will refer to the interacting memory qubit simply as $M$, the system of memory qubits which have previously interacted as $\bar{M}$, and that of those which are yet to interact as $\tilde{M}$. For each such qubit, we define a ``classical basis" $\{\ket{0}, \ket{1}\}$. In what follows, we choose the qubit Pauli-$z$ operator of $D$, $\hat{\sigma}^z_D$, to be diagonal in its energy eigenbasis and that of $M$, $\hat{\sigma}^z_M$ to be diagonal in its classical basis. We are free to choose a phase convention, which we keep fixed, in defining the Pauli-$x$ and -$y$ operators, $\hat{\sigma}^x$ and $\hat{\sigma}^y$ respectively, of these systems. We will also need

\begin{align*}
2\hat{\sigma}^0 =  \hat{\mathds{1}} + \hat{\sigma}^z  & \qquad 2\hat{\sigma}^+ = \hat{\sigma}^x + i \hat{\sigma}^y \\
2\hat{\sigma}^1 =  \hat{\mathds{1}} - \hat{\sigma}^z & \qquad 2\hat{\sigma}^- = \hat{\sigma}^x - i \hat{\sigma}^y \\
\end{align*}

\noindent on these systems as well. Finally, we will use

\begin{align*}
\zeta \equiv \avg{\hat{\sigma}_M^0} - \avg{\hat{\sigma}_M^1}= \avg{\hat{\sigma}_M^z} \\
\end{align*}

\noindent as a shorthand for the population bias of $M$ in its $z$ basis.

The device operates cyclically; $D$ interacts with $M$ for an interaction time $\tau$ before $\mathds{M}$ moves by one site to the right, and the sequence repeats. Each interaction consists of two simultaneous processes, each in contact with a different thermal reservoir. In the first, the demon undergoes \emph{intrinsic transitions}, wherein it exchanges a unit $\Delta$ of energy with a ``hot" reservoir at temperature $T_h$. In the second, the joint system $DM$ undergoes \emph{cooperative transitions}, wherein the demon exchanges $\Delta$ of energy with a ``cold" reservoir at temperature $T_c < T_h$, and the interacting qubit is flipped. Intrinsic transitions occur with rates $\Gamma_{g \rightarrow e}$ and $\Gamma_{g \leftarrow e}$, and cooperative transitions with rates $\Gamma_{g0 \rightarrow e1}$ and $\Gamma_{g0 \leftarrow e1}$. These rates are chosen to satisfy detailed balance

\begin{align*}
\frac{\Gamma_{g \rightarrow e}}{\Gamma_{g \leftarrow e}} = e^{-\beta_h \Delta} && \frac{\Gamma_{g0 \rightarrow e1}}{\Gamma_{g0 \leftarrow e1}} = e^{-\beta_c \Delta} \rm{,} \\
\end{align*}

\noindent where the $\beta_i = 1/T_i$ are the inverse temperatures, and we have chosen units such that the Boltzmann factor is 1. We thus describe each transition by a Lindblad jump operator

\begin{align*}
\hat{L}_{g \rightarrow e} = \sqrt{\Gamma_{g \rightarrow e
}} \hat{\sigma}_D^{-} \otimes \hat{\mathds{1}}_M & \quad \hat{L}_{g0 \rightarrow e1} = \sqrt{\Gamma_{g0 \rightarrow e1}} \hat{\sigma}_D^{-} \otimes \hat{\sigma}^{-}_M \\
\hat{L}_{g \leftarrow e} = \sqrt{\Gamma_{g \leftarrow e}} \hat{\sigma}_D^{+} \otimes \hat{\mathds{1}}_M & \quad \hat{L}_{g0 \leftarrow e1} = \sqrt{\Gamma_{g0 \leftarrow e1}} \hat{\sigma}_D^{+} \otimes \hat{\sigma}^{+}_M \rm{.} \\
\end{align*}

\noindent Finally, we define 

\begin{align*}
\varepsilon \equiv \tanh \left[ \frac{\left(\beta_c - \beta_h \right) \Delta}{2} \right] \\
\end{align*}

\noindent as a parameter quantifying the magnitude of the thermal gradient.

Formally, our interaction sequence is generated by the time-dependent Lindbladian

\begin{align*}
\mathcal{L}_{D \mathds{M}}(t) = \sum_{j = 1}^{\infty} \mathcal{L}^{(j)}_{DM}  \Theta(j \tau - t ) \Theta \left[t - (j - 1)\tau \right] \rm{,} \\
\end{align*}

\noindent where $\mathcal{L}_{DM}^{(j)}$ is the time-independent interaction Lindbladian between $D$ and the $j$th qubit in $\mathds{M}$, described by the aforementioned Lindblad jump operators (see Appendix~\ref{sec:Lindbladian} for details). $\Theta$ is the Heaviside step function, and so these factors ``switch on" the interaction $\mathcal{L}_{DM}^{(j)}$ for $t \in \left[(j - 1)\tau, j\tau \right)$. In the limit of many interactions, the system reaches a periodic steady state $\hat{\rho}_{D\mathds{M}}^{(\rm{ss})}$, for which

\begin{equation}
 \hat{\rho}_{D \mathds{M}}^{\rm{(ss)}} = \hat{\rho}_{D \mathds{M}}\left(n\tau \right) = \hat{\rho}_{D \mathds{M}}\left[(n + 1)\tau\right] \rm{,}
\label{eqn:GenPSS}
\end{equation}

\noindent where 

\begin{align}
\hat{\rho}_{D\mathds{M}}(t) &\equiv  \Phi_{t} \left[\hat{\rho}_{D \mathds{M}}(0)\right] = e^{\int_{0}^{t} \mathcal{L}_{D \mathds{M}}(s) ds} \left[\hat{\rho}_{D \mathds{M}}(0)\right]
\label{eqn:seqevolution} \\ \nonumber
\end{align}

\noindent is the state of the full system at time $t$, and $n \in \mathds{Z}^+$ is a sufficiently large positive integer. In what follows, we will only be interested in the performance of the device over a single interaction in periodic steady state. We therefore omit the explicit time dependence where appropriate and denote quantities corresponding to the outgoing qubit in the interaction as primed and those to the incoming as unprimed \{e.g., $\hat{\rho}_{M} \equiv \hat{\rho}_{M}(n \tau)$ and $\hat{\rho}_{M}^{\prime} \equiv \hat{\rho}_{M}\left[(n + 1) \tau\right]$\}. Similarly, we will refer to the corresponding interaction Lindbladian simply as $\mathcal{L}_{DM}$, dropping the qubit label for convenience.

There are two special cases of this model which are of interest to us. The first is the \emph{classical} case, in which the evolution during each interaction is constrained to population dynamics in the eigenbasis of $\hat{\sigma}_D^z \otimes \hat{\sigma}_M^z$. The interaction $\mathcal{L}_{DM}$ is ``classical" in the sense that if the initial state of $DM$ is diagonal in this basis, then the evolution will remain so throughout. That is to say that the evolution does not mix together populations and coherences in this basis. Additionally, it drives the state of $DM$ to a unique fixed point $\hat{\rho}^{(\rm{fp})}_{DM} = \hat{\rho}^{(\rm{fp})}_{D} \otimes \hat{\rho}^{(\rm{fp})}_{M}$, which is a product of classical states of $D$ and $M$.

The second special case is the \emph{uncorrelated} case, in which the state of $\mathds{M}$ is initially a product, and we neglect any correlations that are built over repeated interactions with the demon. Though the full dynamics, generated by $\mathcal{L}_{D \mathds{M}}(t)$, is a sequence of local quantum operations, it may either build or consume correlations in the full state, since there is a common degree of freedom, $D$, between all of them.  In general, one or both of these special cases may hold.

Figure~\ref{fig:model}(b) shows two cyclic processes over which energy is pumped against the thermal gradient (dashed arrows). The horizontal bit-flip transitions (dot-dashed arrows), which complete each cycle, are provided by the translation of the memory. If the sum of the probabilities of these processes is greater than that for the reverse processes, then there is a net flow of energy against the gradient, and we say the system is \emph{refrigerating}. Note that both processes flip the qubit from the $\ket{0}$ state to the $\ket{1}$ state, so a record of the refrigeration is left on the memory. We thus define

\begin{align*}
Q_{h \rightarrow c} \equiv \frac{\Delta}{2} \left(\zeta^{\prime} - \zeta \right) \\
\end{align*}

\noindent as the heat flow from the hot reservoir to the cold reservoir during the interaction, which is negative when the system is refrigerating. Note that this property of the model removes any ambiguity in our definition of work. We also define

\begin{align*}
\Delta S_{M} \equiv S(\hat{\rho}^{\prime}_M) - S(\hat{\rho}_M) \rm{,} \\
\end{align*}

\noindent  where

\begin{align*}
S(\hat{\rho}) \equiv -\tr \left( \hat{\rho} \log \hat{\rho} \right) \\
\end{align*}

\noindent is the von Neumann entropy. The change $\Delta S_{M}$ represents the entropy dumped onto the memory. When $\Delta S_{M} < 0$, the memory is being \emph{erased}. 

In the uncorrelated classical regime, our model reduces to that of Ref.~\cite{mandal2012maxwell}, where it was first introduced. There, it was shown that

\begin{equation}
Q_{h \rightarrow c}(\beta_c - \beta_h) + \Delta S_{M} \geq 0\rm{,}
\label{eqn:LocalClausius}
\end{equation}

\noindent \\ \noindent in this regime. We refer to Eq.~(\ref{eqn:LocalClausius}) as the \emph{local Clausius inequality}. It represents a strict trade-off between the refrigeration ($Q_{h \rightarrow c} < 0$) and erasure ($\Delta S_{M} < 0$) phases of the device. This trade-off can be understood geometrically. In the absence of correlations or quantum coherence, the interacting qubit is driven monotonically along the $z$ axis of the Bloch sphere to its fixed point, $\hat{\rho}^{\mathrm{(fp)}}_M$, which has positive bias $\zeta^{\mathrm{(fp)}} = \varepsilon > 0$. Erasure and refrigeration correspond to approaching this fixed point from $\zeta < \varepsilon$ and $\zeta > \varepsilon$, respectively. So long as this evolution is monotonic, the trade-off is necessarily strict.

It is noted in Ref.~\cite{mandal2012maxwell} that when correlations cannot be neglected, this result generalizes to

\begin{align}
Q_{h \rightarrow c}(\beta_c - \beta_h) + \Delta S_{M} \geq \Delta I_{D:M} \label{eqn:genLocalClausius} \rm{,} \\ \nonumber
\end{align}

\noindent where 

\begin{align*}
\Delta I_{A:B} = I(A:B)^{\prime} - I(A:B) \\
\end{align*}

\noindent is the change in the quantum mutual information

\begin{align*}
I(A:B) \equiv S(\hat{\rho}_A) + S(\hat{\rho}_B) - S(\hat{\rho}_{AB}) \rm{.} \\
\end{align*}

\noindent $\Delta I_{D:M}$ can be either positive or negative in general, so the trade-off in Eq.~(\ref{eqn:LocalClausius}) becomes no longer strict. This suggests the possibility of a third thermodynamically nontrivial phase, in which the device is refrigerating and erasing simultaneously. We expect this to occur when correlations between $D$ and $M$ induce nonmonotonic evolution on $M$ alone. However, Eq.~(\ref{eqn:genLocalClausius}) is inconvenient in that it does not involve only terms that can be calculated from the evolution on the memory alone and so requires knowledge of the periodic steady state, Eq.~(\ref{eqn:GenPSS}). One of our contributions is to provide a constraint free of this limitation, which is also strict in the presence of correlations and makes their utility as a thermodynamic resource manifest.

\section{Main Results}
\label{sec:Main}

\subsection{Global Clausius inequality}
\label{sec:gcinequality}

Our first result is to recover the second law in terms of the relevant thermodynamic variables in the fully quantum correlated case. This is the \emph{global Clausius inequality}, which we state as a theorem.

\begin{theorem}
  \label{thm:GlobalClausius}
  The global Clausius inequality,
\begin{equation}
Q_{h \rightarrow c} (\beta_c - \beta_h) + \Delta S_{M} - \Delta I_{M:\tilde{M}} \geq 0, \\
\label{eqn:GlobalClausius}
\end{equation}
which represents an information-work trade-off of the thermodynamic quantities defined above, holds for a periodic steady state in the fully quantum correlated regime. Accordingly, simultaneous refrigeration and erasure is possible only if it is attended by the consumption of correlations.
\end{theorem}

\noindent Equation~(\ref{eqn:GlobalClausius}) represents a strict three-way trade-off among refrigeration, erasure, and the generation of correlations $\Delta I_{M:\tilde{M}}$, all of which may be calculated from knowledge of the initial and final states of the memory alone. The minus sign on $\Delta I_{M:\tilde{M}}$ places consumption of correlations as a resource; the more negative this term, the more negative the local terms can be.

To provide a sketch of the proof of Theorem~\ref{thm:GlobalClausius} (full details can be found in Appendix \ref{sec:gcProof}), we first note

\begin{align}
Q_{h \rightarrow c} (\beta_c - \beta_h) + \Delta S_{M \tilde{M}} \geq \Delta I_{D:M\tilde{M}} \rm{,} \label{eqn:gcinequality1} \\ \nonumber
\end{align}

\noindent which follows in the same manner as Eq.~(\ref{eqn:genLocalClausius}) from monotonic evolution to the fixed point $\hat{\rho}^{\mathrm{(fp)}}_{DM}$, but including the system $\tilde{M}$ as an ancilla. Next, we have

\begin{align*}
\Delta I_{D : M \tilde{M}} &= I(D : M \tilde{M})^{\prime} - I(D : M \tilde{M}) \\
& \geq I(D : \tilde{M})^{\prime} - I(D : M \tilde{M}) \\
& = I(D : M \tilde{M}) - I(D : M \tilde{M}) = 0 \rm{.} \\
\end{align*}

\noindent The first inequality follows from strong subaddivity: Mutual information is nonincreasing under partial trace. The last step follows from the periodic steady-state condition, Eq.~(\ref{eqn:GenPSS}). That is, the mutual information between $D$ and the qubits yet to interact---$M \tilde{M}$ before the interaction and $\tilde{M}$ after the interaction---is a constant across interactions in steady state. Thus

\begin{align*}
Q_{h \rightarrow c} (\beta_c - \beta_h) + \Delta S_{M \tilde{M}} \geq 0 \rm{,} \\
\end{align*}

\noindent and using

\begin{align*}
\Delta S_{M \tilde{M}} = \Delta S_{M} + \Delta S_{\tilde{M}} -  \Delta I_{M:\tilde{M}} \rm{,} \\
\end{align*}

\noindent where $\Delta S_{\tilde{M}} = 0$ due to the fact that $\tilde{M}$ does not participate in the interaction, therefore gives the result. 

We see that simultaneous refrigeration and local erasure is not forbidden by Eq.~(\ref{eqn:GlobalClausius}) so long as correlations are consumed, but it remains to be seen that our model actually exhibits such behavior. We address this in the following sections.

\subsection{Matrix product density operator solution}
\label{sec:MPDOsoln}

The matrix product state formalism has been very successful as an efficient representation of the quantum correlations present in a one-dimensional spin-chain \cite{fannes1992finitely, vidal2003efficient, perez2006matrix}. Because dynamics in contact with a thermal bath are dissipative, we need a generalization of this formalism to mixed states and classical correlations. This is the \emph{matrix product density operator} (MPDO) formalism, introduced in Refs.~\cite{verstraete2004MPDO, zwolak2004mixed} and further developed in Refs.~\cite{bonnes2014superoperators, werner2014positive, cui2015variational}. The MPDO description of a mixed state of $N$ $d$-dimensional spins with periodic boundary conditions is given by

\begin{equation}
\label{eqn:MPDOdef}
   \hat{\rho}_{\rm{MPDO}} = \sum_{\vec{i}} \tr \left(\mathcal{P} \prod_{j = 1}^{N} \hat{A}^{i_j} \right) \bigotimes_{j = 1}^N \hat{\sigma}^{i_j} \rm{,}
\end{equation}

\noindent where the $i_{j} \in \{0, 1, +, -\}$, and the $\hat{A}^{i_j}$ are $\chi_{j - 1} \times \chi_{j}$ matrices corresponding to the spin at site $j$.  $\mathcal{P}$ is the path-ordering operator, which places operators of higher $j$ to the left in the product. It was shown in Ref.~\cite{verstraete2004MPDO} that the description Eq.~(\ref{eqn:MPDOdef}) can be obtained from a corresponding pure matrix product state by tracing out ancillary degrees of freedom on each of the spins. It thus reduces to those for pure matrix product states and classically correlated distributions as special cases. Because we are interested in the periodic steady-state behavior of our model, we restrict ourselves to the single-site \emph{translationally invariant} case where the $\hat{A}^{i_j}$ are independent of the site label $j$. We thus have that the $\chi_j = \chi$ are all equal. This quantity is known as the \emph{bond dimension} of the MPDO, which quantifies the degree to which the state is correlated. Note that, in this representation, the state is specified by $d^2 \chi^2$ complex parameters, an exponential improvement --- when bond dimension is polynomial in $N$ --- over the $O(d^N)$ complex parameters required to specify each of the matrix elements of $\hat{\rho}_{\rm{MPDO}}$ individually. 

In our model, we let $\hat{\rho}_{\mathds{M}}(0) = \hat{\rho}_{\rm{MPDO}}$ be the parametrization of the initial quantum correlated state of the memory $\mathds{M}$, which is uncorrelated with the initial state of $D$. We first solve the Lindblad master equation for an individual interaction

\begin{align*}
\frac{d \hat{\rho}_{DM}}{dt} &= \mathcal{L}_{DM} \left(\hat{\rho}_{DM}\right) \\
\end{align*}

\noindent analytically in MATHEMATICA, obtaining an expression for the quantum operation

\begin{align*}
\phi_{\tau} &\equiv e^{\mathcal{L}_{DM}\tau} \rm{.} \\ 
\end{align*}

\begin{figure}
\subfloat[]{\label{fig:circuitmpo}
  \includegraphics[width=0.5\textwidth]{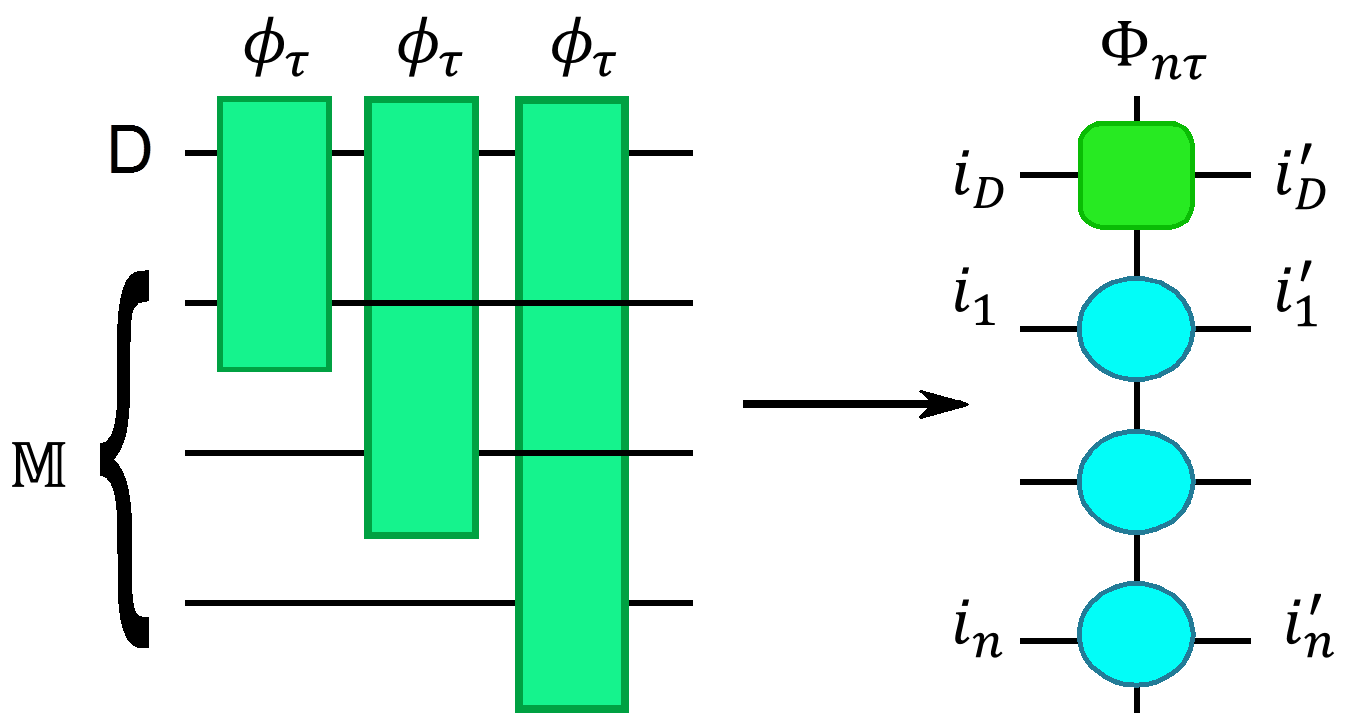}
 	}\\
\subfloat[]{\label{fig:circuitcomp}
  \includegraphics[width=0.5\textwidth]{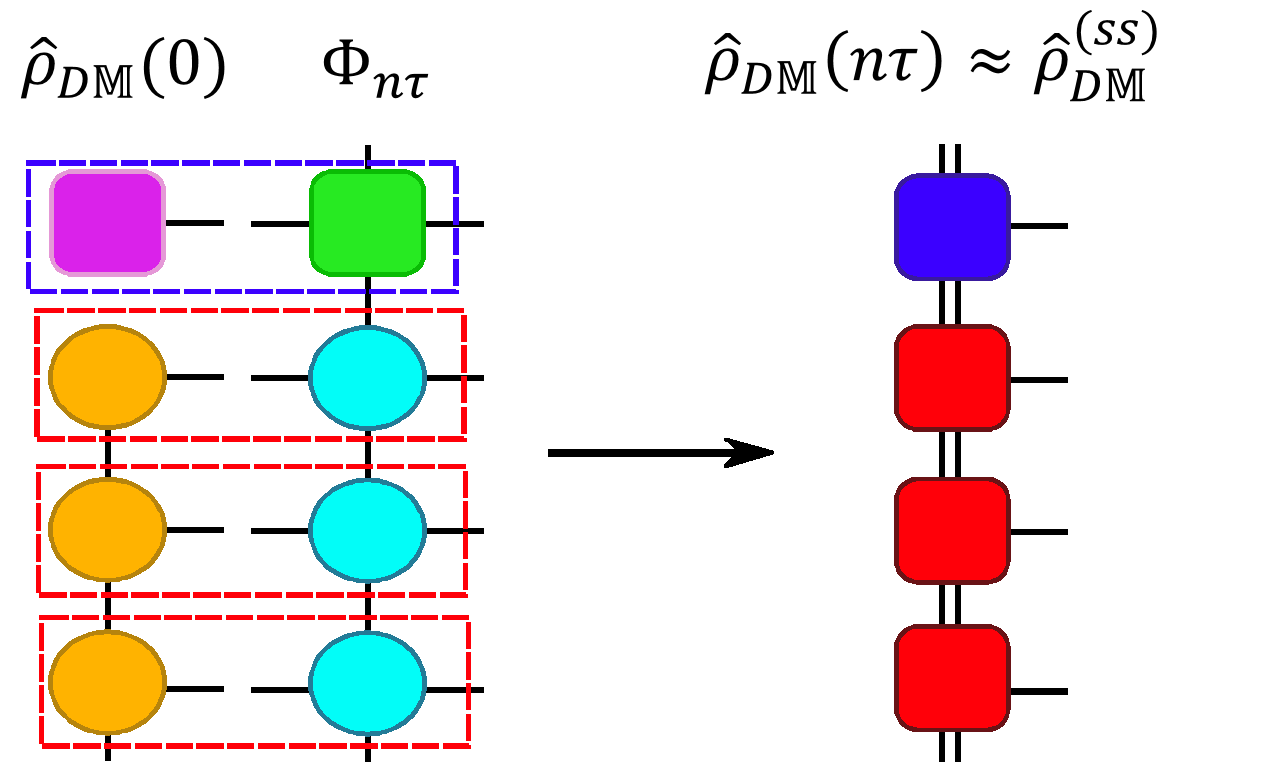}
 	}
\caption{(Color online) A pictorial representation of our time evolution method, which relies on a matrix product density operator (MPDO) description, for $n = 3$. (a) On the left is a schematic depiction of our interaction sequence, with quantum operations vectorized to let us represent the sequence as a quantum circuit. The shared degree of freedom is the demon qubit $D$, and the remaining degrees of freedom are the memory $\mathds{M}$. We indicate that a particular qubit is not acted upon by a given quantum operation by drawing the qubit's line \emph{over} the quantum operation's box. We parametrize the sequence as a matrix product operator (MPO), shown on the right, with physical indices labeled according to the corresponding qubit degrees of freedom. (b) We multiply our initial MPDO, for which the state of $\mathds{M}$ is translationally invariant and uncorrelated with that of $D$, by the MPO describing our interaction sequence to obtain an approximation to the steady state. This is done by performing a sequence of index contractions, shown on the left as dotted boxes. The approximation to the steady state, which is valid for sufficiently large $n$, is shown as the resulting MPDO on the right. All boundary conditions are taken to be periodic.}
\label{fig:MPDOschematic}
\end{figure}

\noindent We then update our initial state $\hat{\rho}_{D \mathds{M}}(0)$ to $\hat{\rho}_{D \mathds{M}}(n \tau) \approx \hat{\rho}_{D \mathds{M}}^{\mathrm{(ss)}}$ according to the sequential evolution, Eq.~(\ref{eqn:seqevolution}), using the method schematically depicted in Fig.~\ref{fig:MPDOschematic}. Here we have vectorized the Hilbert space, representing quantum operations as matrices and density matrices as vectors (see Ref.~\cite{zwolak2004mixed} for details). In the figure, we represent each on-site tensor $\hat{A}^{i_j}$ by a shape (box or circle) and that tensor's indices (either $i_j$ or the matrix indices of $\hat{A}^{i_j}$ for a particular value of $i_j$) as lines from the corresponding shape. Tensors with a contracted index between them are joined by a line, and uncontracted indices are represented by open lines. This pictorial scheme is thus a convenient representation for tensor contractions.

In Fig.~\ref{fig:MPDOschematic}(a), we apply a generalization of the result from Ref.~\cite{schoen2005sequential} to parametrize our series of sequential quantum operations as a matrix product operator (MPO). Figure~\ref{fig:MPDOschematic}(b) shows the update, which is performed by simply multiplying the initial MPDO, $\hat{\rho}_{D \mathds{M}}(0)$, by the MPO, $\Phi_{n \tau}$. This scheme permits us to only store $3\times 4 \chi^2$ parameters: the tensor corresponding to $M \tilde{M}$, that to $\bar{M}$, and a boundary tensor to $D$. The full details are given in Appendix \ref{sec:MPDOupdate}.

We next apply the classicality of our interaction, $\phi_{\tau}$, to simplify our MPDO description for $\hat{\rho}^{\rm{(ss)}}_{D M \tilde{M}}$ in terms of operationally defined quantities as our second main result.

\begin{theorem}
The periodic steady state of $D M \tilde{M}$ prior to the interaction on $M$ may be expressed as

\begin{equation}
\hat{\rho}^{\rm{(ss)}}_{D M \tilde{M}} = \sum_{\vec{k}} p_{\vec{k}} \hat{\rho}^{(\vec{k})}_D \otimes \hat{\rho}^{(\vec{k})}_{M \tilde{M}} \rm{,}
\label{eqn:cqseparablemain}
\end{equation}

\noindent where the $\vec{k} \in \{0, 1\}^n$ are classical bit strings of length $n$, which correspond to classical records on the memory subsystem $\bar{M}$. $p_{\vec{k}}$ is the probability of the record $\vec{k}$ ($\sum_{\vec{k}} p_{\vec{k}} = 1$), and $\hat{\rho}^{(\vec{k})}_{M \tilde{M}}$ is the reduced state of $M \tilde{M}$ conditioned upon that record. $\hat{\rho}_{D}^{(\vec{k})}$ is the reduced state of $D$ upon interacting with the string of uncorrelated classical pure qubits specified by $\vec{k}$, which is classical. 
\label{thm:cqseparable}
\end{theorem}

\noindent The proof and formal expressions for $p_{\vec{k}}$, $\hat{\rho}_{D}^{(\vec{k})}$, and $\hat{\rho}^{(\vec{k})}_{M \tilde{M}}$ can be found in Appendix \ref{sec:MPDOupdate}. Theorem \ref{thm:cqseparable} says that the steady state is always separable and given by a convex combination over all possible classical histories of the memory. Note that the memory part $\hat{\rho}^{(\vec{k})}_{M \tilde{M}}$ could contain quantum coherence. It is convenient in that it expresses the steady state $\hat{\rho}_{D M \tilde{M}}^{\rm{(ss)}}$ in terms of the input (the $p_{\vec{k}}$ and the $\hat{\rho}_{M \tilde{M}}^{(\vec{k})}$) and the interaction (the $\hat{\rho}_{D}^{(\vec{k})}$) separately. This will allow us to engineer states of $\mathds{M}$ to achieve a desired thermodynamic behavior. It is also a useful analytic tool when it is tractable (i.e., there are relatively few terms in the sum). We further demonstrate its utility by using it to prove that the demon's performance in steady state is invariant under local phase rotations on the memory, which we state formally in the following corollary:

\begin{corollary}
The Clausius terms, $Q_{h \rightarrow c}$, $\Delta S_{M}$, and $\Delta I_{M:\tilde{M}}$, take the same values for the input states $\hat{\rho}_{\mathds{M}}(0)$ and $\mathcal{U}_\mathds{M}^{z, \varphi}\left[\hat{\rho}_{\mathds{M}}(0) \right]$ in steady state, where 

\begin{align*}
\mathcal{U}_\mathds{M}^{z, \varphi} \left(\hat{\rho}_{\mathds{M}}\right) = \left(e^{-i (\varphi/2) \hat{\sigma}^z}\right)^{\otimes n} \hat{\rho}_{\mathds{M}} \left(e^{i (\varphi/2) \hat{\sigma}^z}\right)^{\otimes n}
\end{align*}

\noindent constitutes an individual rotation about the $z$ axis applied transversally to every qubit in $\mathds{M}$.
\label{cor:cqseparable1}
\end{corollary}

To see this, we first note that we calculate $\hat{\rho}^{\rm{(ss)}}_{D M \tilde{M}}$ for $\mathcal{U}_\mathds{M}^{z, \varphi}\left[\hat{\rho}_{\mathds{M}}(0) \right]$ by making the replacement 

\begin{align*}
\hat{\rho}_{M \tilde{M}}^{(\vec{k})} \mapsto \mathcal{U}^{z, \varphi}_{M \tilde{M}} \left(\hat{\rho}_{M \tilde{M}}^{(\vec{k})} \right)
\end{align*}

\noindent in Eq.~(\ref{eqn:cqseparablemain}). This is because transversal phase rotations affect neither the probabilities of classical measurements $p_{\vec{k}}$ nor the states $\hat{\rho}^{(\vec{k})}_D$ by their definition. Furthermore, the states $\hat{\rho}_D^{(\vec{k})}$ are classical and so invariant under phase rotation. We thus have

\begin{align*}
\sum_{\vec{k}} p_{\vec{k}} \hat{\rho}^{(\vec{k})}_D \otimes  \mathcal{U}^{z, \varphi}_{M \tilde{M}} \left(\hat{\rho}_{M \tilde{M}}^{(\vec{k})} \right) &= \sum_{\vec{k}} p_{\vec{k}} \ \mathcal{U}^{z, \varphi}_{D M \tilde{M}} \left( \hat{\rho}^{(\vec{k})}_D \otimes \hat{\rho}_{M \tilde{M}}^{(\vec{k})} \right)\\ 
&= \mathcal{U}^{z, \varphi}_{D M \tilde{M}} \left(\hat{\rho}_{D M \tilde{M}}\right) \rm{.}
\end{align*}

\noindent That is

\begin{align}
(\mathcal{I}_{D} \otimes \mathcal{U}_M^{z, \varphi})(\hat{\rho}^{\rm{(ss)}}_{DM}) = \mathcal{U}_{DM}^{z, \varphi}(\hat{\rho}^{\rm{(ss)}}_{DM}) \rm{,} \label{eqn:rotinvariance} \\ \nonumber
\end{align}

\noindent where $\mathcal{I}$ is the identity superoperator. We see from Eq.~(\ref{eqn:rotinvariance}) that $\mathcal{L}_{DM}$ commutes with $\mathcal{U}^{z, \varphi}_{M}$ on $\hat{\rho}^{\rm{(ss)}}_{DM}$, as the cooperative transition term commutes with $\mathcal{U}_{DM}^{z, \varphi}$, and the intrinsic transition and Hamiltonian terms commute with $\mathcal{U}_{M}^{z, \varphi}$ (Appendix \ref{sec:Lindbladian}). Corollary~\ref{cor:cqseparable1} therefore follows from the fact that all of the terms in Eq.~(\ref{eqn:GlobalClausius}) are invariant under transversal phase rotations on $\hat{\rho}_{M \tilde{M}}$ and $\hat{\rho}_{M \tilde{M}}^{\prime}$.

\subsection{Simultaneous refrigeration and erasure}
\label{sec:Numerics}

We first examine a special case, which is simple enough that the calculation of Eq.~(\ref{eqn:cqseparablemain}) is analytically tractable but also correlated enough to reveal nontrivial thermodynamic behavior.

\begin{figure}
\subfloat[]{\label{fig:classicalghzphase}
  \includegraphics[width=0.45\textwidth]{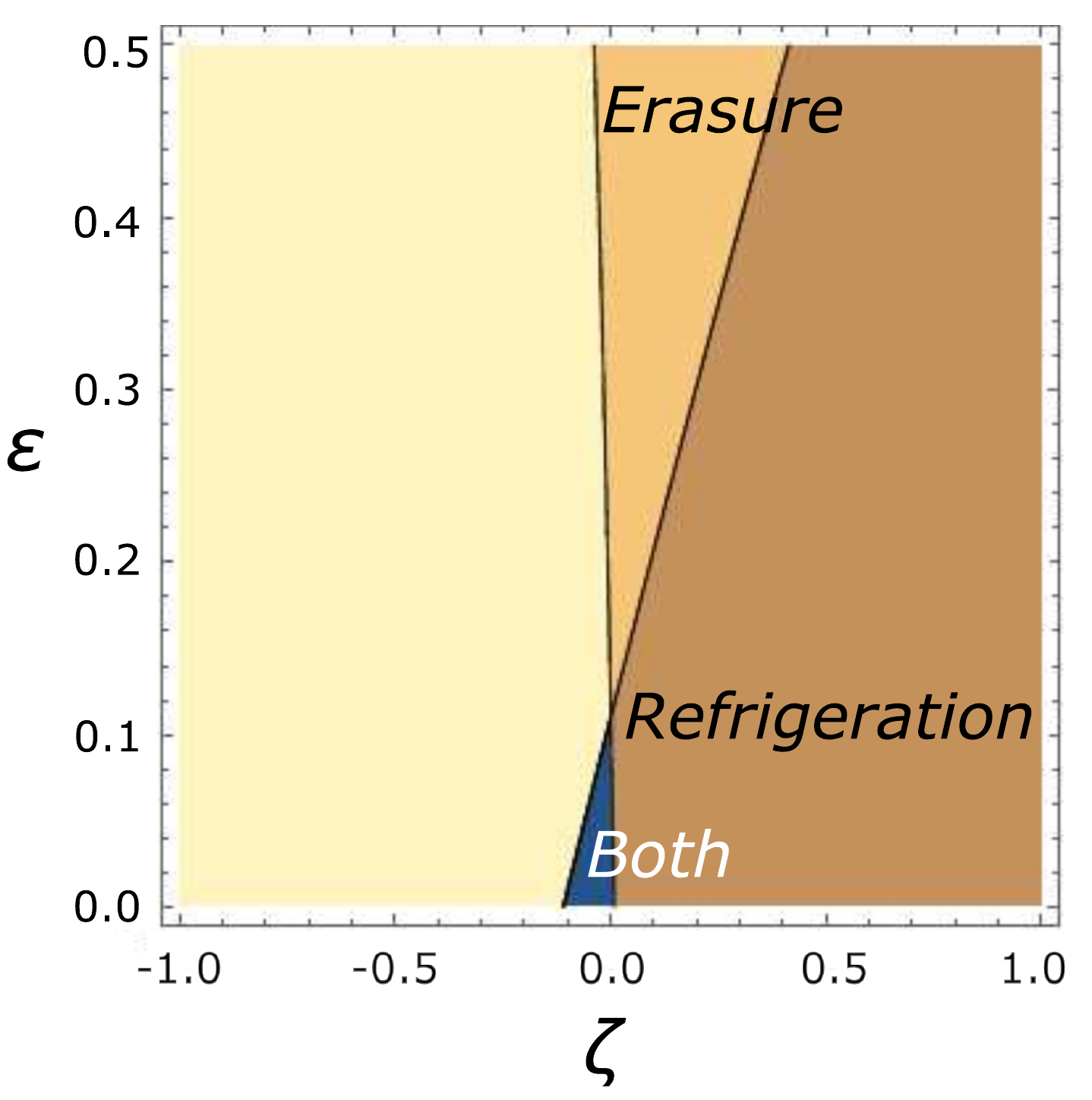}
 	}\\
\subfloat[]{\label{fig:ghzvtau}
  \includegraphics[width=0.45\textwidth]{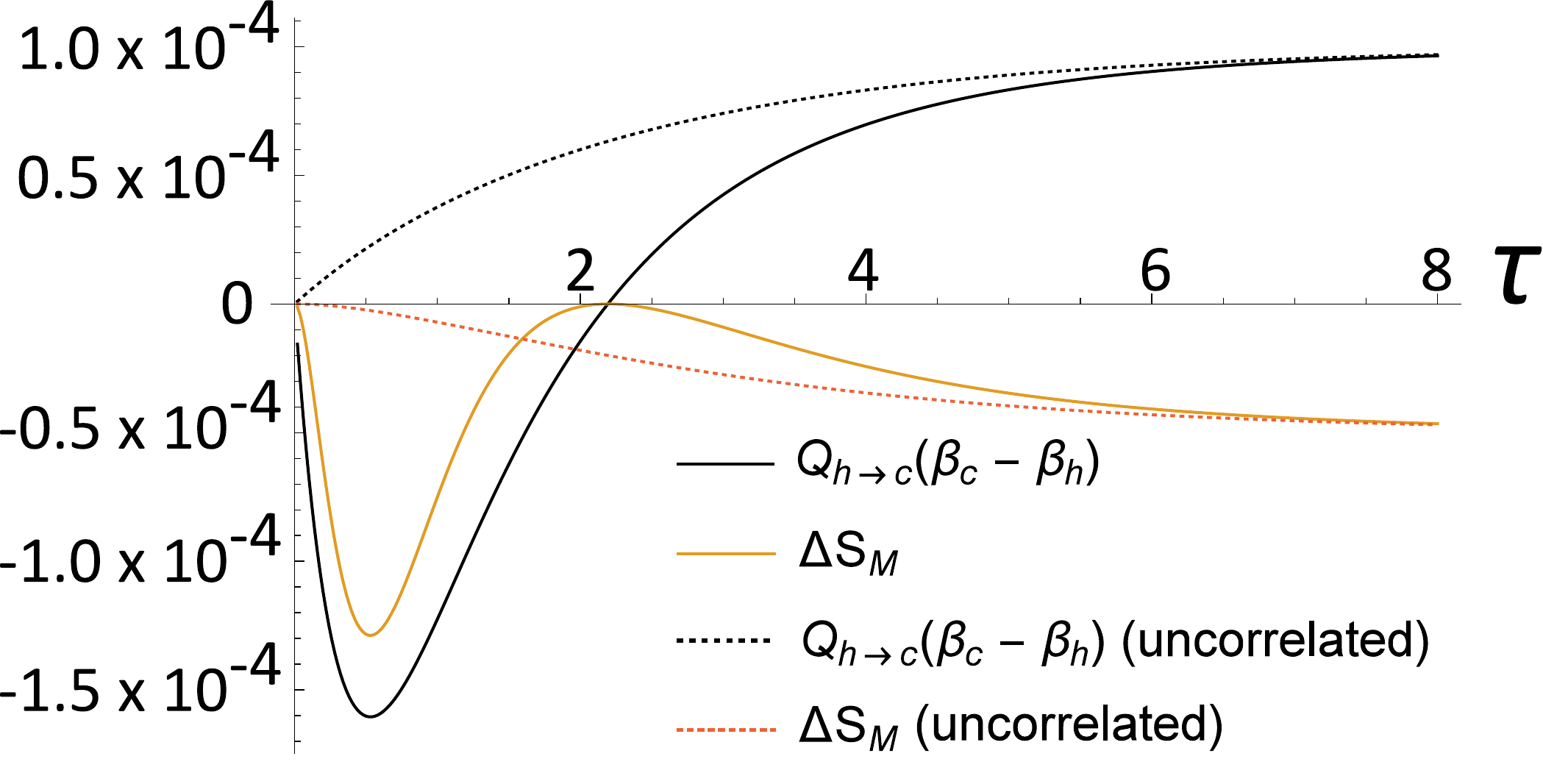}
 	}
\caption{(Color online) (a) The nonequilibrium phase diagram for input states from the GHZ-correlated family described in Observation \ref{obs:newphase}, analogous to Fig.~2 in Ref.~\cite{mandal2012maxwell}, for $\tau = 0.3$ (units of $1/\Delta$ for $\hbar = 1$). For $\zeta = \pm 1$, the input is a product, and it is entangled for $\zeta \in \left(-1, 1\right)$, with the strongest correlations at $\zeta = 0$, for which it is the GHZ state. Note that, for every state in this family, the reduced state of any qubit is diagonal in the $z$ basis. We see that correlations shift the phase boundaries to induce a region of simultaneous refrigeration and erasure near the uncorrelated phase transition triple point ($\varepsilon = \zeta = 0$). (b) Dependence of the Clausius terms, $Q_{h \rightarrow c}(\beta_c - \beta_h)$ and $\Delta S_M$, on the interaction time, $\tau$ when $\varepsilon = 0.01$ in comparing the GHZ state for $\zeta = 0$---solid black and orange (gray) curves, respectively---to the product of its single-qubit reduced states, which are all maximally mixed---dotted black and orange (gray) curves, respectively. The reduced evolution of the interacting memory qubit is monotonic when there is a strict trade-off of refrigeration and erasure; both cannot be simultaneously negative. This is obeyed by the uncorrelated input, but the correlated input shows a departure from monotonicity for $\tau \lesssim 2.2$. For long-enough interaction times, the effect of correlations is ``washed out" as the terms for correlated input approach the values for their uncorrelated-input counterparts.}
\label{fig:newphase}
\end{figure}

\begin{observation}
Consider the family of Greenberger-Horne-Zeilinger- (GHZ) correlated states parametrized by a bias $\zeta$,

\begin{equation}
\label{eqn:ghzfamily}
\ket{\psi}_{\mathds{M}} = \frac{1}{\sqrt{2}} \left(\sqrt{1 + \zeta}\ket{0}^{\otimes N} + \sqrt{1 - \zeta}\ket{1}^{\otimes N} \right) \rm{,}
\end{equation}

\noindent with $N$ sufficiently large to allow the system to reach a periodic steady state.
There exists a nonequilibrium phase of simultaneous refrigeration ($Q_{h \rightarrow c} < 0$) and erasure of memory ($\Delta S_{M} < 0$), when the memory is initially prepared to be the quantum entangled state $\ket{\psi}_{\mathds{M}}$ with a bias $\zeta$ and a temperature gradient $\varepsilon$ in the region illustrated in Fig.~\ref{fig:newphase}(a).
\label{obs:newphase}
\end{observation}

\noindent We see that the reduced state on each qubit for any state in the family Eq.~(\ref{eqn:ghzfamily}) is \emph{locally classical} with bias $\zeta$ and that for $\zeta = \pm 1$, $\ket{\psi}_{\mathds{M}}$ is a product. For $\zeta \in (-1, 1)$, however, the state is entangled, with maximal entanglement at $\zeta = 0$. 

For this family, we calculate the Clausius terms in Eq.~(\ref{eqn:GlobalClausius}) analytically by letting the system undergo $n$ interactions --- so it reaches periodic steady state --- and tracing out $\bar{M}$. In Eq.~(\ref{eqn:cqseparablemain}), only two classical histories appear in the sum, and we have

\begin{align}
\hat{\rho}_{M \tilde{M}} = \sum_{k \in \{0, 1\}} \left[\frac{1 + (-1)^k \zeta}{2} \right] \left(\hat{\sigma}^k \right)^{\otimes (N - n)} \rm{,} \label{eqn:GHZssInput} \\ \nonumber
\end{align}

\noindent and

\begin{align}
\hat{\rho}^{\prime}_{M \tilde{M}} = \sum_{k \in \{0, 1\}} \left[\frac{1 + (-1)^k \zeta}{2} \right] \hat{\rho}^{\prime (k)}_{M} \otimes \left(\hat{\sigma}^k \right)^{\otimes (N - n - 1)} \rm{,} \label{eqn:GHZssInput} \\ \nonumber
\end{align}

\noindent where

\begin{align*}
\hat{\rho}_M^{\prime (k)} = \tr_{D \bar{M}} \left\{\Phi_{(n + 1) \tau} \left[\hat{\rho}_{D}(0) \otimes \hat{\sigma}_{M}^{k} \otimes \left(\hat{\sigma}^{k}\right)^{\otimes n}_{\bar{M}} \right]\right\} \\
\end{align*}

\noindent is the state of $M$ following its interaction for an uncorrelated tape of pure qubits in state $\ket{k}$, in periodic steady state. The global entropy change $\Delta S_{M \tilde{M}}$ then takes the particularly simple form

\begin{align*}
\Delta S_{M \tilde{M}} = \frac{1}{2} \left[(1 + \zeta) S(\hat{\rho}^{\prime (0)}_{M}) + (1 - \zeta) S(\hat{\rho}^{\prime (1)}_{M}) \right]\rm{,}
\end{align*}

\noindent Note that this quantity is always non-negative. By comparison, the local quantities

\begin{align*}
\Delta S_{M} &= S\left[\left(\frac{1 + \zeta}{2}\right) \hat{\rho}^{\prime (0)}_{M} + \left(\frac{1 - \zeta}{2}\right) \hat{\rho}^{\prime (1)}_{M} \right] \\
&\quad - S\left[\frac{1}{2} \left( \hat{\mathds{1}} + \zeta \hat{\sigma}^z \right) \right]
\end{align*}

\noindent and

\begin{align*}
Q_{h \rightarrow c}  = \frac{\Delta}{2}\left[\left(\frac{1 + \zeta}{2}\right) \zeta^{\prime (0)} + \left(\frac{1 - \zeta}{2}\right) \zeta^{\prime (1)} - \zeta \right] \rm{,}
\end{align*}

\noindent where $\zeta^{\prime (i)} = \tr\left(\hat{\sigma}_M^{z} \hat{\rho}_M^{\prime (i)} \right)$, may be either positive or negative. Using the expressions for the $\zeta^{\prime (i)}$ given in Ref.~\cite{mandal2012maxwell}, we analytically construct an analogous nonequilibrium phase diagram, Fig.~\ref{fig:newphase}(a), for $\tau = 0.3$ (in units of $1/\Delta$, where we have set $\hbar = 1$). We see that the effect of the correlation is to introduce a region of simultaneous refrigeration and erasure near the uncorrelated phase transition triple point at $\varepsilon = \zeta = 0$. $\zeta$ thus decreases below zero over the interaction in this region, indicating a deviation from monotonicity in the reduced evolution of $M$. This is the primary role of correlations in our thermodynamic model: in periodic steady state, initial correlations in $\mathds{M}$ induce classical correlations between $D$ and $M$ prior to their interaction, allowing $\hat{\rho}^{M}$ to pass through hitherto inaccessible regions of the Hilbert space in its evolution. 

\begin{figure}
 \includegraphics[width=0.5\textwidth]{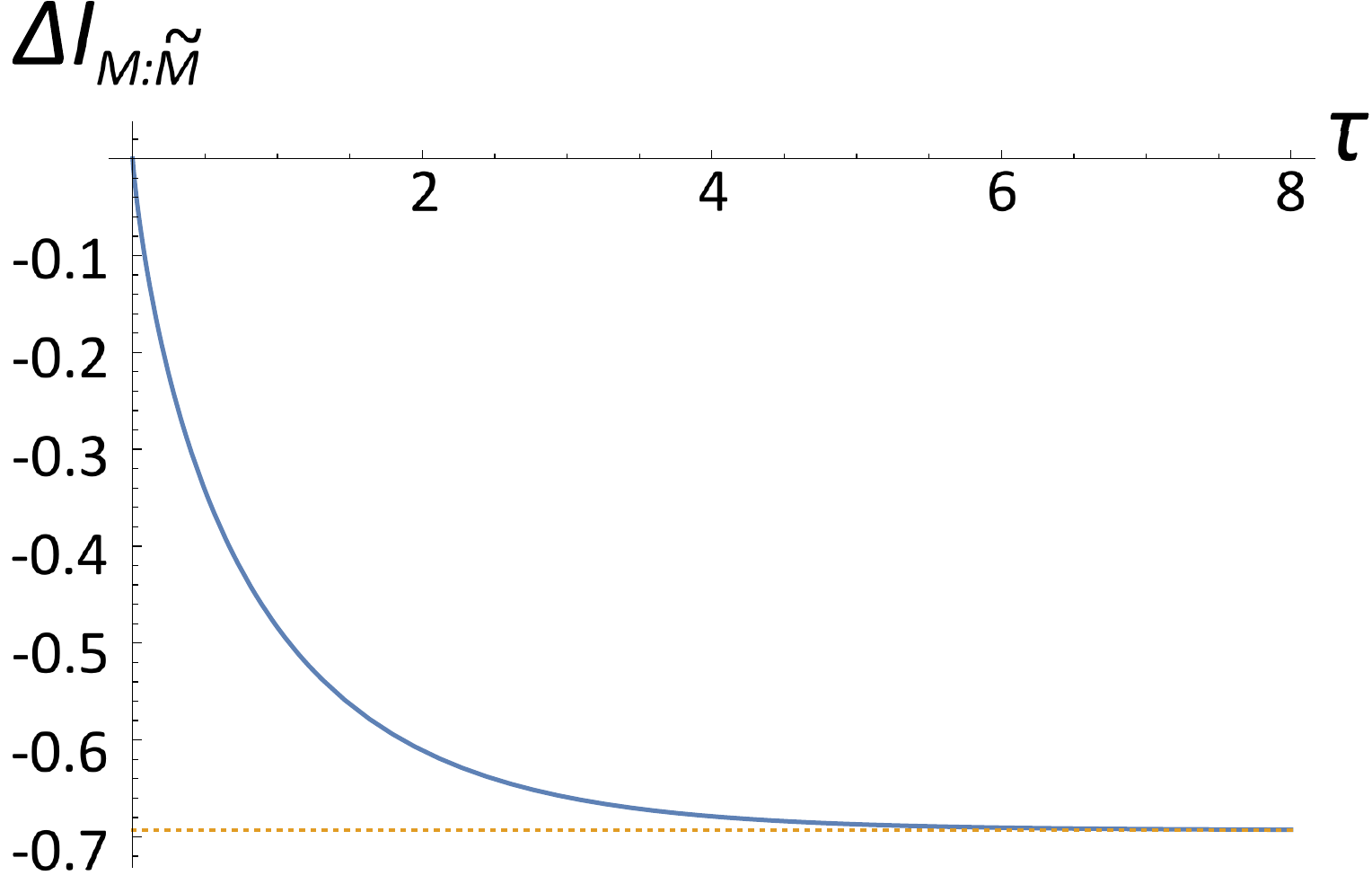}
	\caption{(Color online) Generation of correlations $\Delta I_{M:\tilde{M}}$ between the interacting qubit $M$ and the qubits $\tilde{M}$ yet to interact, as a function of the interaction time $\tau$, for the GHZ input of Fig.~\ref{fig:newphase}(b). $\tau$ is in units of $1/\Delta$ for $\hbar = 1$, and $\varepsilon = 0.01$. Correlations are always being consumed, and so this term offsets the others in the Clausius inequality such that simultaneous refrigeration and erasure does not violate the second law for this input family. The dotted line is at $\Delta I_{M:\tilde{M}} = -\ln 2$ and is meant to provide a visual aid, as it represents the maximal possible consumption of correlations for this state.}
  \label{fig:ghzdivtau}
\end{figure}

We examine this more closely in Fig.~\ref{fig:newphase}(b), where we plot the quantities in the local Clausius inequality, Eq.~(\ref{eqn:LocalClausius}), for the input given in Eq.~(\ref{eqn:ghzfamily}) as a function of $\tau$ for $\varepsilon = 0.01$ and $\zeta = 0$ as compared to the corresponding terms for product input with the same reduced state on every qubit. In Fig.~\ref{fig:ghzdivtau}, we plot the generation of correlations $\Delta I_{M:\tilde{M}} = \Delta S_{M} - \Delta S_{M \tilde{M}}$. We first note that, though correlations are always being consumed, the demon is not always able to use this to its advantage. This is as we expect, since the behavior must approach that of the uncorrelated case as $\tau \rightarrow \infty$. For $\tau \lesssim 0.54$, we see increasing simultaneous refrigeration and erasure with additional interaction time. Past $\tau \approx  0.54$, additional interaction time does not afford better thermodynamic performance, and, eventually, we see that correlations actually begin to hinder the demon's erasure at $\tau \approx 1.6$. Finally, we see that the demon's behavior approaches that of the uncorrelated case as $\tau$ becomes large, as we expect.

We see from the diagonal form of $\hat{\rho}_{M \tilde{M}}$ that we would have had the same effect if the input had been prepared in the equivalent classically correlated (i.e., diagonal in the computational basis) state 

\begin{align*}
\hat{\rho}^{\rm{(c)}}_{\mathds{M}} = \frac{1}{2} \left[ (1 + \zeta) \left(\hat{\sigma}^0 \right)^{\otimes N} + (1 - \zeta)\left(\hat{\sigma}^1 \right)^{\otimes N} \right] \rm{.}
\end{align*}

\noindent We are then led to ask: Is there anything to be said for the role of quantumness in this model? To address this, we consider another example, which will give our final result.

\subsection{Quantum thermodynamic advantage}
\label{sec:quantadv}

As a final result, we observe that correlations can enable our model to exploit quantum coherence to gain a thermodynamic advantage

\begin{observation}
Using an arbitrary orthonormal $\vec{n}$ basis, 
\begin{align*}
\ket{+\vec{n}} &= \cos \left(\theta/2 \right) \ket{0} + e^{i \phi} \sin \left(\theta/2 \right) \ket{1} \rm{,} \\
\ket{-\vec{n}} &= \sin \left(\theta/2 \right) \ket{0} - e^{i \phi} \cos \left(\theta/2 \right) \ket{1} \rm{,} \\
 \end{align*}

\noindent 
with $\theta \in [0, \pi ]$, and $\phi \in [0, 2\pi)$, we define two states: (i)
the GHZ-correlated input family in an $\vec{n}$ basis $\hat{\rho}_{\mathds{M}}^{(\rm{q})} = \ketbra{\psi}{\psi}_{\mathds{M}}$ such that 

\begin{equation}
\label{eqn:rotghzfamily}
\ket{\psi}_{\mathds{M}} = \frac{1}{\sqrt{2}} \left(\sqrt{1 + \zeta_{\vec{n}}}\ket{+\vec{n}}^{\otimes N} + \sqrt{1 - \zeta_{\vec{n}}}\ket{-\vec{n}}^{\otimes N} \right) \rm{,}
\end{equation}

\noindent with $N$ sufficiently large to allow the system to reach steady state, and (ii) its corresponding classically-correlated family,
\begin{equation}
\hat{\rho}^{(\rm{c})}_{\mathds{M}} = \sum_{\vec{k}} \left(\bigotimes_{j = 1}^N \hat{\sigma}^{k_j} \right) \hat{\rho}_{\mathds{M}}^{(\rm{q})} \left(\bigotimes_{j = 1}^N \hat{\sigma}^{k_j} \right)
\end{equation}

\noindent with the $k_j \in \{0, 1\}$, obtained by eliminating all off-diagonal matrix elements from $\hat{\rho}_{\mathds{M}}^{(\rm{q})}$ in the $z$ basis. The quantum entangled state $\hat{\rho}_{\mathds{M}}^{(\rm{q})}$ is advantageous in memory erasure over the classically correlated mixed state $\hat{\rho}^{(\rm{c})}_{\mathds{M}}$, whose bias for every single-qubit reduced state is identical to that of $\hat{\rho}_{\mathds{M}}^{(\rm{q})}$.
\label{obs:coherence}
\end{observation}

\noindent This comparison might correspond to the scenario where the experimenter only has the ability to prepare correlations in a fixed basis, versus that where the experimenter has this ability in addition to the ability to perform phase rotations in an orthogonal basis. Thanks to Corollary \ref{cor:cqseparable1}, we know that this second ability is the only addition over the first needed to observe the full range of thermodynamic performance. Because the states $\hat{\rho}_{\mathds{M}}^{(\rm{q})}$ and $\hat{\rho}_{\mathds{M}}^{(\rm{c})}$ allow for all possible classical histories with some probability, the demon's performance cannot be easily calculated with Eq.~(\ref{eqn:cqseparablemain}) as with Eq.~(\ref{eqn:ghzfamily}). We thus obtain our results numerically using the MPDO description.

\begin{figure}
\subfloat[]{\label{fig:3dadvregion}
  \includegraphics[width=0.45\textwidth]{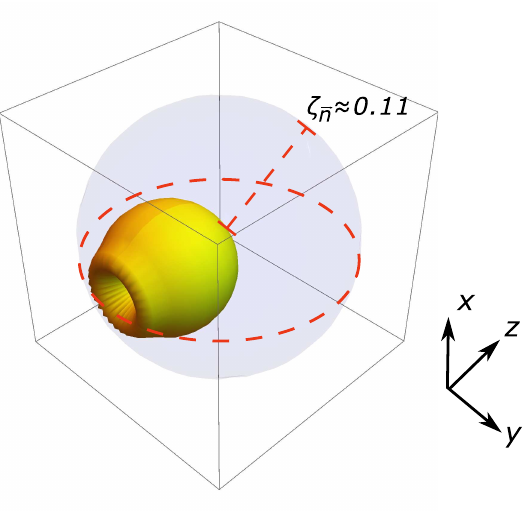}
 	}\\
\subfloat[]{\label{fig:roterasure}
  \includegraphics[width=0.45\textwidth]{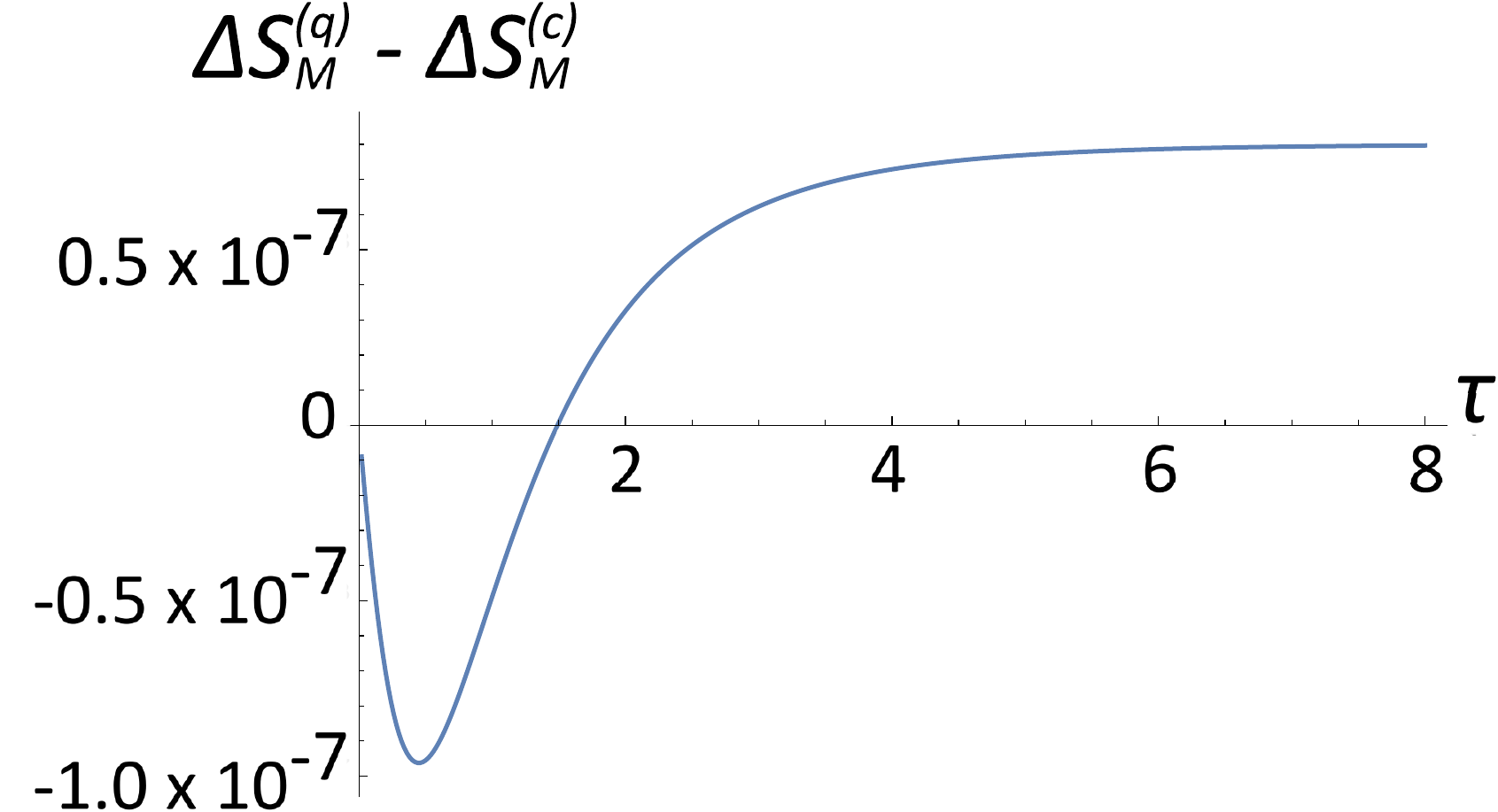}
 	}
\caption{(Color online) (a) The three-dimensional surface whose enclosure is the set of local Bloch vectors for which the locally coherent GHZ-correlated family attains a quantum erasure advantage over its corresponding classically correlated family, whose off-diagonal elements in the $z$ basis have been projected out. Here $\tau = 0.3$ and $\varepsilon = 0.01$. We see clearly the predicted azimuthal symmetry about the $z$ axis, and the advantage vanishes along this axis as we expect, since the two inputs are the same in this case. (b) The difference between the erasure performance for the quantum correlated locally coherent input state and that of its decohered counterpart for $\varepsilon = 0.01$, $\phi = \pi/2$, $\theta = 0.01$, and $\zeta_{\vec{n}} = -0.02$ as a function of the interaction time $\tau$. A quantum advantage is demonstrated when this quantity is negative. We see that the advantage is several orders of magnitude smaller and lost much faster than the advantage afforded by classical correlations of Fig.~\ref{fig:newphase}(b).}
\label{fig:quantumadv}
\end{figure} 

In Fig.~\ref{fig:quantumadv}(a), we plot the exact region for which the advantage exists [i.e., $\Delta S^{\rm{(q)}} < \Delta S^{\rm{(c)}}$ and $\Delta S^{\rm{(q)}} < 0$]. We see here clearly the azimuthal symmetry of Corollary \ref{cor:cqseparable1} and that this advantage sharply disappears as $\theta \rightarrow 0$. In Fig.~\ref{fig:quantumadv}(b) for $\varepsilon = 0.01$, we show the difference in erasures between the quantum and classical inputs as a function of $\tau$. The quantum coherent state $\hat{\rho}_{\mathds{M}}^{(\rm{q})}$ has an advantage when this quantity is negative. This plot bears some resemblance to Fig.~\ref{fig:newphase}(b). This is no coincidence. In the same way as classical correlations in $DM$ can allow the Bloch vector of $M$ to momentarily move away from its fixed point along the $z$ axis, they can also allow it to do so in the $x$-$y$ plane, though the effect is smaller by several orders of magnitude. This deviation from monotonicity appears as an erasure advantage since our interaction $\phi_{\tau}$ treats the classical and coherent components of the evolution separately. The refrigerative performances of the two inputs are thus identical as we expect.

What is remarkable is that even though both $\hat{\rho}_{\mathds{M}}^{(\rm{q})}$ and $\hat{\rho}_{\mathds{M}}^{(\rm{c})}$ share the same classical histories in Eq.~(\ref{eqn:cqseparablemain}), the small amount of coherence on the reduced state $\hat{\rho}^{\rm{(q)}}_M$ seems to be able to seed the production of more coherence. This is especially surprising given the fact that our dynamics $\phi_{\tau}$ also has many classical features. The explanation is that the coherences on the conditional states of $M$ for each classical history \emph{interfere} such that there is more coherence on $\hat{\rho}_{M}$ after the evolution than before.

\section{Conclusion}
\label{sec:discussion}

We have shown that our model of an autonomous Maxwell demon can utilize correlations in its memory to gain a thermodynamic advantage over its uncorrelated counterpart. That is, as predicted by the global Clausius inequality, it performs simultaneous refrigeration and erasure, which is impossible without correlations, and exploits quantum coherence to enhance the performance of memory erasure. This work represents a proof-of-principle, and we hope that it paves the way to invent a quantum thermodynamic device which is more well suited to utilizing correlations.

There are two concluding discussions. First, in this work, entanglement did not demonstrate a genuine advantage over classical correlations. This is because of the long-range nature of GHZ correlations. Namely, tracing out a distant subsystem gives a classically correlated state but does not affect the thermodynamic performance, which is determined by the local interaction between the demon qubit and one of memory qubits. However, as far as we have explored, some translationally invariant, short-range-correlated states (such as the one-dimensional cluster state \cite{briegel2001persistent}) did not exhibit the simultaneous refrigeration-erasure phase. More demanding numerical calculation seems to be needed to generalize beyond these simple initial states, to address the question about strict supremacy of entanglement.

Second, the simplicity of our model could bring the testing of quantum coherence as a thermodynamic resource closer to experimental feasibility, since the model shares similarities with quantum heat engines \cite{scully2003extracting, geva2010irreversible, linden2010small, cleuren2012cooling, esposito2012stochastically, gelbwaserklimovsky2013minimal, venturelli2013minimal, boukobza2013breaking, correa2013performance, correa2014quantum, alicki2014quantum, leggio2015quantum}. Though we consider a concrete experimental proposal to be beyond the current scope of the paper, we note that there are similarities between our model and the quantum heat engines considered in Refs.~\cite{vogl2009laser, vogl2011collisional, szczygielski2013markovian}. Here, a two-level system (TLS), such as a pair of electronic energy levels of an optically active rubidium atom, is periodically driven by a red-detuned laser field and independently coupled to two thermal baths, such as the photon bath of a cavity and the ``dephasing bath" of a buffer gas, at different temperatures. We envision the TLS serving the role of the demon in these systems, while the laser realizes the memory. As a buffer gas molecule approaches the rubidium atom, it perturbs the energy gap of the TLS into resonance with the laser, allowing the atom to become excited. Once the buffer molecule has scattered, and the gap of the TLS has widened back to its unperturbed value, the rubidium atom may fluoresce, radiating a quantum of energy into the photon bath. The authors of Ref.~\cite{szczygielski2013markovian} note the existence of a regime where the dephasing bath may be cooled at the expense of work provided by the laser, with energy transfer to the photon bath mediated by the TLS. This regime has been observed experimentally as reported in Refs.~\cite{vogl2009laser, vogl2011collisional}. Though we do not expect our model's Hamiltionian to be isomorphic to that of this system, we do expect that it can be treated using our formalism in the case where the laser field state is correlated.

The most severe imperfections to this setup would come from limitations on the experimenter's ability to produce a precisely correlated state. We refrain from speculation on what experimental techniques are needed to prepare such states. Here we found that the novel thermodynamic behavior was rather sensitive to the strength of the correlations [in the phase diagram, Fig.~\ref{fig:newphase}(a), the simultaneous-refrigeration-erasure phase occupies a narrow range of $\zeta$], but not so sensitive to classical correlations versus entanglement. This latter feature, should it survive in other architectures, could be promising for prospective experiments. Finally, our investigation was by no means exhaustive, so it remains to be seen what types of correlation are needed to optimize the thermodynamic performance of these particular experimental setups. It would be interesting to see if the utility of correlations as a thermodynamic resource survives in these architectures as well.

\begin{acknowledgments}
The work was supported in part by National Science Foundation Grants No. PHY-1212445 and No. PHY-1521016.
\end{acknowledgments}

\appendix
\section{Exposition of the Interaction Lindbladian}
\label{sec:Lindbladian}

Here we give the details of our interaction Lindbladian $\mathcal{L}_{DM}$. As mentioned in the main text, the intrinsic and cooperative transitions are characterized by the Lindblad jump operators

\begin{align*}
\hat{L}_{g \rightarrow e} &= \sqrt{\Gamma_{g \rightarrow e}} \hat{\sigma}_D^- \otimes \hat{\mathds{1}}_{M} \\
\hat{L}_{g \leftarrow e} &= \sqrt{\Gamma_{g \leftarrow e}} \hat{\sigma}_D^+ \otimes \hat{\mathds{1}}_{M} \\
\end{align*}

\noindent for the intrinsic transitions, and

\begin{align*}
\hat{L}_{g0 \rightarrow e1} &= \sqrt{\Gamma_{g0 \rightarrow e1}} \hat{\sigma}_D^- \otimes \hat{\sigma}_M^- \\
\hat{L}_{g0 \leftarrow e1} &= \sqrt{\Gamma_{g0 \leftarrow e1}} \hat{\sigma}_D^+ \otimes \hat{\sigma}_M^+ \\
\end{align*}

\noindent for the cooperative transitions. Here, $\hat{\sigma}^{\pm} = \frac{1}{2} \left(\hat{\sigma}^x \pm i \hat{\sigma}^y \right)$ are the qubit ladder operators, and $\hat{\sigma}^{\{x, y\}}$ are qubit Pauli-$x$ and -$y$ operators. Once again, the transition rates are chosen to satisfy detailed balance,

\begin{align*}
\frac{\Gamma_{g \rightarrow e}}{\Gamma_{g \leftarrow e}} = e^{-\beta_h \Delta} && \frac{\Gamma_{g0 \rightarrow e1}}{\Gamma_{g0 \leftarrow e1}} = e^{-\beta_c \Delta} \rm{.} \\
\end{align*}

\noindent Finally, the demon has an intrinsic Hamiltonian,

\begin{align*}
\hat{H}_D = \Delta \ket{e}\bra{e} = \frac{\Delta}{2} \left(\hat{\mathds{1}}_D - \hat{\sigma}^{z}_D \right) \mathrm{,} \\
\end{align*}

\noindent and so the full dynamics of the interaction are generated by the Lindbladian

\begin{equation}
\label{eqn:Lindblad}
\mathcal{L}_{DM} \left(\hat{\rho}_{DM}\right) = -i [\hat{H}_D \otimes \hat{\mathds{1}}_M, \hat{\rho}_{DM}] + \sum_{i} \mathcal{L}_i \left(\hat{\rho}_{DM} \right)\rm{,}
\end{equation}

\noindent where

\begin{align*}
\mathcal{L}_i \left(\hat{\rho}_{DM}\right) &= \hat{L}_i \hat{\rho}_{DM} \hat{L}^{\dagger}_i - \frac{1}{2} \{ \hat{L}^{\dagger}_i \hat{L}_i, \hat{\rho}_{DM}  \} \mathrm{,} \\
\end{align*}

\noindent is the term corresponding to a particular jump operator. $[.,.]$ and $\{.,.\}$ are the commutator and anticommutator, respectively, and we have chosen units such that $\hbar = 1$. As mentioned in the main text, $\mathcal{L}_{g0 \rightarrow e1}$ and $\mathcal{L}_{g0 \leftarrow e1}$ commute with $\mathcal{U}_{DM}^{z, \varphi}$ as defined in Corollary \ref{cor:cqseparable1}, due to the fact that $\hat{\sigma}_D^z \otimes \hat{\sigma}_M^z$ commutes with $\hat{\sigma}_D^i \otimes \hat{\sigma}_M^j$ for $i, j \in \{x, y\}$. Clearly $\mathcal{L}_{g \rightarrow e}$, $\mathcal{L}_{g \leftarrow e}$, and the Hamiltonian terms commute with $\mathcal{U}_{M}^{z, \varphi}$.

$\mathcal{L}_{DM}$ has a fixed point $\hat{\rho}_{DM}^{(\rm{fp})}$, for which

\begin{align*}
\mathcal{L}_{DM} \left(\hat{\rho}_{DM}^{(\rm{fp})} \right) = \hat{\rho}_{DM}^{(\rm{fp})} \\
\end{align*}

\noindent and given by a product

\begin{align*}
\hat{\rho}_{DM}^{(\rm{fp})} = \hat{\rho}_{D}^{(\rm{fp})} \otimes \hat{\rho}_{M}^{(\rm{fp})} \mathrm{,} \\
\end{align*}

\noindent where
 
\[
\hat{\rho}_{D}^{(fp)} = \frac{1}{1 + e^{\beta_{h} \Delta}} \left( \begin{array}{cc}
1 & 0 \\
0 &  e^{\beta_{h} \Delta}
\end{array} \right)
\] \\

\noindent in the energy eigenbasis, and

\[
\hat{\rho}_{M}^{(fp)} = \frac{1}{1 + e^{(\beta_h - \beta_c)\Delta}} \left( \begin{array}{cc}
1 & 0 \\
0 &  e^{(\beta_h - \beta_c)\Delta}
\end{array} \right)
\] \\

\noindent in the $z$ basis. The form of this fixed point will be important for the following proofs.

\section{Generalized Clausius Inequality}
\label{sec:gcProof}

Here we give a proof of Eqs.~(\ref{eqn:LocalClausius}), (\ref{eqn:genLocalClausius}), and (\ref{eqn:gcinequality1}). As stated in the main text, we have monotonic evolution to the fixed point in a single interaction $(\phi_{\tau} \otimes \mathcal{I}_{\tilde{M}})$

\begin{equation}
 D(\hat{\rho}_{D M \tilde{M}} ||\hat{\rho}_{DM}^{(\rm{fp})} \otimes \hat{\rho}_{\tilde{M}}) - D(\hat{\rho}^{\prime}_{D M \tilde{M}}||\hat{\rho}_{DM}^{(\rm{fp})} \otimes \hat{\rho}_{\tilde{M}}) \geq 0
\label{eqn:Monotonicity}
\end{equation}

\noindent where
	
\begin{align*}
D(\hat{\rho}||\hat{\sigma}) = -S(\hat{\rho}) - \tr\left(\hat{\rho} \ln \hat{\sigma} \right) \\
\end{align*}

\noindent is the quantum relative entropy. Using

\begin{align*}
\ln \left(\hat{\rho}_A \otimes \hat{\rho}_B\right) &= \ln \hat{\rho}_A \otimes \hat{\mathds{1}}_B  + \hat{\mathds{1}}_A \otimes \ln \hat{\rho}_B \\
\tr\left[\hat{\rho}_{AB} \left(\hat{O}_A \otimes \hat{\mathds{1}}_B \right) \right] &= \tr \left( \hat{\rho}_{A} \hat{O}_A \right) \equiv \avg{\hat{O}_A} \\
\end{align*}

\noindent and

\begin{align*}
\tr \left[ \left(\hat{\rho}^{\prime}_D - \hat{\rho}_D \right) \ln \hat{\rho}_D^{(\rm{fp})} \right] &= 0 \rm{,} \\
\end{align*}

\noindent from the periodic steady-state condition on $D$, together with the fact that the interaction acts as the identity on $\tilde{M}$, Eq.~(\ref{eqn:Monotonicity}) reduces to

\begin{equation}
\Delta S_{D M \tilde{M}} + \tr \left[ \left(\hat{\rho}_M^{\prime} - \hat{\rho}_M \right) \ln \hat{\rho}_M^{(\rm{fp})} \right] \geq 0
\label{eqn:Monotonicity2}
\end{equation}

\noindent Next, we note that

\begin{align*}
\avg{\ln \hat{\rho}_M^{(\rm{fp})}} &= \frac{\Delta}{2}(\beta_h - \beta_c) \left(1 - \zeta \right) - \ln \left[1 + e^{(\beta_h - \beta_c)\Delta} \right] \\
\end{align*}

\noindent All of the constant terms cancel in the difference, and so we are left with

\begin{align*}
\tr \left[ \left(\hat{\rho}_M^{\prime} - \hat{\rho}_M \right) \ln \hat{\rho}_M^{(fp)} \right] = \frac{\Delta}{2} (\zeta^{\prime} - \zeta)(\beta_c - \beta_h)
\end{align*}

\noindent Defining $Q_{h \rightarrow c} \equiv \frac{\Delta}{2}  (\zeta^{\prime} - \zeta)$, Eq.~(\ref{eqn:Monotonicity2}) gives

\begin{equation}
Q_{h \rightarrow c}(\beta_c - \beta_h) + \Delta S_{D M \tilde{M}} \geq 0
\label{eqn:Monotonicity3}
\end{equation}

\noindent Finally, we use the definition of quantum mutual information

\begin{align*}
S(\hat{\rho}_{D M \tilde{M}}) &= S(\hat{\rho}_{D}) + S(\hat{\rho}_{M \tilde{M}}) - I(D:M\tilde{M})
\end{align*}

\noindent and the fact that $\Delta S_{D} = 0$, again from the periodic steady state condition on $D$, in Eq.~(\ref{eqn:Monotonicity3}) to obtain Eq.~(\ref{eqn:gcinequality1})

\begin{align*}
Q_{h \rightarrow c}(\beta_c - \beta_h) + \Delta S_{M \tilde{M}} \geq \Delta I_{D:M\tilde{M}} \\
\end{align*}

\noindent Equation (\ref{eqn:genLocalClausius}) follows in the same manner, but without including $\tilde{M}$, and Eq.~(\ref{eqn:LocalClausius}) requires neglecting correlations ($\Delta I_{D:M} \approx 0$).

\section{Matrix Product Density Operator Update Method}
\label{sec:MPDOupdate}

We describe our matrix product quantum operation shown in Fig.~\ref{fig:MPDOschematic}(a) by its action on the product basis elements in Eq.~(\ref{eqn:MPDOdef}). Using a generalization of the result in Ref.~\cite{schoen2005sequential} from vectors to operators as in Ref.~\cite{zwolak2004mixed}, we update each element by $n$ sequential interactions, $\Phi_{n\tau}$, as
	
\begin{align*}
\Phi_{n\tau} \left(\hat{\sigma}_D^{i_D} \bigotimes_{j = 1}^n \hat{\sigma}^{i_j} \right) = \sum_{i_D^{\prime}, \vec{i}^{\prime}} \tr \left(\hat{B}^{i_D} \mathcal{P} \prod_{j = 1}^n \hat{C}^{i_j^{\prime} i_j} \right) \hat{\sigma}_D^{i_D^{\prime}} \bigotimes_{j = 1}^n \hat{\sigma}^{i_j^{\prime}} \rm{,}
\end{align*}

\noindent where

\begin{align*}
\left(\hat{C}^{i^{\prime} i} \right)_{\alpha \beta} &= \tr{\left[\left(\hat{\sigma}_D^{\alpha} \otimes \hat{\sigma}_M^{i^{\prime}} \right)^{\dagger} \phi_{\tau} \left(\hat{\sigma}_D^{\beta} \otimes \hat{\sigma}_M^{i} \right) \right]} \\
\left(\hat{B}^{i_D} \right)_{\alpha \beta} &= \tr{\left[\left(\hat{\sigma}_D^{\alpha} \right)^{\dagger} \hat{\rho}_{D}(0) \right]} \delta_{i_{D} \beta} \rm{,} \\
\end{align*}

\noindent where $\delta_{i_{D} \beta}$ is the Kronecker delta and the $\hat{\sigma}^{i_D}$ are the basis elements for $D$. These expressions follow from completeness under the \emph{trace inner product} for this operator basis,

\begin{align*}
\avg{\hat{A}, \hat{B}}_{\tr} &= \tr{\hat{A}^{\dagger} \hat{B}} \\
\hat{A} &= \sum_{i} \avg{\hat{\sigma}^i, \hat{A}}_{\tr} \hat{\sigma}^i \\ 
\end{align*}

Thus, after rearranging,

\begin{align}
\hat{\rho}_{D\mathds{M}}(n \tau) &= \sum_{i_D, \vec{i}} \tr \Big[ \left(\hat{B}^{i_D} \otimes \hat{\mathds{1}} \right) \mathcal{P} \prod_{j = n + 1}^{N} \left(\hat{\mathds{1}} \otimes
\hat{A}^{i_j} \right) \nonumber \\ & \quad \times \mathcal{P} \prod_{j = 1}^{n} \left(\sum_{k} \hat{C}^{i_j k} \otimes \hat{A}^{k} \right) \Big] \hat{\sigma}_D^{i_D}  \bigotimes_{j = 1}^N \hat{\sigma}^{i_j} \rm{,} \label{eqn:genpss} \\ \nonumber
\end{align}

\noindent and in the next interaction, we make the replacement

\begin{align*}
\hat{\mathds{1}} \otimes \hat{A}^{i_{n + 1}} \mapsto \sum_{k} \hat{C}^{i_{n + 1} k} \otimes \hat{A}^{k} \rm{.} \\
\end{align*}

\noindent We see that this leaves the description of Eq.~(\ref{eqn:genpss}) approximately unchanged for sufficiently large $N$, $n$. This is the aforementioned periodic steady state of the system.

Finally, we trace out all previously interacted degrees of freedom to obtain

\begin{align}
\hat{\rho}_{D M \tilde{M}} &= \sum_{i_D, \vec{i}} \tr \Big[ \left(\hat{B}^{i_D} \otimes \hat{\mathds{1}} \right) \mathcal{P} \prod_{j = n + 1}^{N} \left(\hat{\mathds{1}} \otimes
\hat{A}^{i_j} \right) \nonumber \\ & \quad \times \left(\sum_{k} \hat{D}^k \otimes \hat{A}^{k} \right)^n \Big] \hat{\sigma}^{i_D}  \bigotimes_{j = n + 1}^N \hat{\sigma}^{i_j} \rm{,} \label{eqn:updatedMPDO} \\ \nonumber
\end{align}

\noindent where

\begin{align*}
\hat{D}^k = \sum_{l} \hat{C}^{l k} \tr{\hat{\sigma}^l} \rm{.} \\
\end{align*}

\noindent After the next interaction, we have

\begin{align}
\hat{\rho}^{\prime}_{D M \tilde{M}} &= \sum_{i_D, \vec{i}} \tr \Big[ \left(\hat{B}^{i_D} \otimes \hat{\mathds{1}} \right) \mathcal{P} \prod_{j = n + 2}^{N} \left(\mathds{1} \otimes
\hat{A}^{i_j} \right) \nonumber \\ & \quad \times \left(\sum_{k} \hat{C}^{i_{n + 1} k} \otimes \hat{A}^{i_{n + 1}} \right) \left(\sum_{k} \hat{D}^k \otimes \hat{A}^{k} \right)^n \Big] \label{eqn:updatedMPDOprime} \\ & \quad \times \hat{\sigma}^{i_D}  \bigotimes_{j = n + 1}^N \hat{\sigma}^{i_j} \rm{.} \nonumber \\ \nonumber
\end{align} 

\noindent We see from this expression that its convergence with $n$ depends on that of the transfer matrix $\left(\sum_{k} \hat{D}^k \otimes \hat{A}^{k}\right)$ to its fixed point. It is therefore exponentially convergent at the rate of the inverse correlation length of the outgoing MPDO, as we expect.

We now simplify Eq.~(\ref{eqn:updatedMPDO}) using is the classicality of our interaction, $\phi_{\tau}$, which implies $\hat{D}^{\pm} = 0$. Rearranging, and making the identifications 

\begin{align*}
\hat{\rho}^{(\vec{k})}_D &=  \sum_{i_D} \tr \left[ \hat{B}^{i_D} \mathcal{P} \left( \prod_{j = 1}^{n} \hat{D}^{k_j} \right) \right] \hat{\sigma}_D^{i_D} \rm{,} \\
p_{\vec{k}} \hat{\rho}^{(\vec{k})}_{M \tilde{M}} &= \sum_{\vec{i}} \tr \left[\mathcal{P} \left(\prod_{j = n + 1}^{N} \hat{A}^{i_j} \right) \left(\prod_{j = 1}^{n} \hat{A}^{k_j} \right) \right] \bigotimes_{j = n + 1}^N \hat{\sigma}^{i_j} \rm{,} \\
\end{align*}

\noindent for $\vec{k} \in \{0, 1\}^{n}$ gives our Theorem~\ref{thm:cqseparable},

\begin{align}
\hat{\rho}_{D M \tilde{M}} &= \sum_{\vec{k}} p_{\vec{k}} \hat{\rho}^{(\vec{k})}_D \otimes \hat{\rho}^{(\vec{k})}_{M \tilde{M}} \label{eqn:cqseparable} \rm{,} \nonumber
\end{align}

\noindent where

\begin{align*}
\hat{\rho}^{(\vec{k})}_D &= \mathcal{P} \prod_{j = 1}^{n} \mathcal{T}^{k_j} \left[\hat{\rho}_{D}(0) \right] \rm{,} \\
\hat{\rho}^{(\vec{k})}_{M \tilde{M}} &= \frac{1}{p_{\vec{k}}} \tr_{\bar{M}} \left[\left( \hat{\mathds{1}}_{M \tilde M} \bigotimes_{j = 1}^n \hat{\sigma}^{k_j} \right) \hat{\rho}_{\mathds{M}}(0) \right] \rm{,} \\
\end{align*}

\noindent and

\begin{align*}
\mathcal{T}^{k_j} \left(\hat{\rho}_D\right) &\equiv \tr_{M} \phi_{\tau} \left(\hat{\rho}_D \otimes \hat{\sigma}^{k_j}_{M} \right) \rm{,} \\
p_{\vec{k}} &= \tr \left[\left( \hat{\mathds{1}}_{M \tilde M} \bigotimes_{j = 1}^n \hat{\sigma}^{k_j} \right) \hat{\rho}_{\mathds{M}}(0) \right] \rm{.} \\
\end{align*}

The classicality of the $\hat{\rho}^{(\vec{k})}_D$ follows from the uniqueness of the fixed point of the transfer operator $\mathcal{T}^{k}$, which is the periodic steady state of the demon upon interacting with a string of uncorrelated pure classical qubits, $\ket{k}^{\otimes{n}}$, and found in Ref.~\cite{mandal2012maxwell}. Because this fixed point is classical and unique, $\mathcal{T}^{k}$ must be dephasing in the energy eigenbasis, and so the $\hat{\rho}^{(\vec{k})}_D$ are classical as well.

\end{document}